\documentstyle[preprint,aps]{revtex}
\begin{document}
\draft
\preprint{}

\title{{\it Ab Initio} Linear Response Study of SrTiO$_{3}$}
\author{Chris LaSota, Cheng-Zhang Wang, Rici Yu, and Henry Krakauer} 
\address{Department of Physics,
  College of William and Mary, Williamsburg, Virginia 23187--8795}
\date{March 5, 1996}
\maketitle
\begin{abstract}
  The lattice instabilities of perovskite structure oxides are
  responsible for many of their interesting properties, such as
  temperature-dependent ferroelectric phase transitions.
  First-principles calculations using linear response theory provide
  an accurate means to determine the lattice dynamics throughout the
  entire Brillouin zone (BZ). Using the LAPW linear response
  method\cite{yu94}, we have previously carried out such a study on
  ferroelectric KNbO$_3$\cite{yu95}.  We present here the
  results of a similar investigation for cubic SrTiO$_3$.
  While KNbO$_3$ has only ferroelectric-type instabilities,
  SrTiO$_3$ exhibits both ferroelectric (FE) and
  antiferrodistortive (AFD) instabilities.  We correctly
  predict the known instability at the R-point in
  the Brillouin zone, which is responsible for the 
  AFD phase transition to the tetragonal structure at
  about 105~K.  Furthermore, the phase space of the
  ferroelectric instability is greatly reduced compared to KNbO$_3$.
  Antiferrodistortive instabilities exist in one-dimensional
  cylindrical tubes extending along the entire R-M-R line in the
  Brillouin zone.  The essentially 1-D character of these
  tubes corresponds to real-space planar instabilities
  characterized by rotations of oxygen octahedra.
\end{abstract}

\narrowtext

\section{Introduction}

Strontium titanate has been the subject of ongoing theoretical
and experimental investigation for over 30 years,
due to its unusual properties.  In cubic
perovskite structure KNbO$_3$ and BaTiO$_3$,
transitions to ferroelectric phases occur as the
temperature is lowered.  SrTiO$_3$, which has the
same high-temperature structure, behaves somewhat
differently in that it undergoes an
antiferrodistortive (AFD) transition as the
temperature drops below about 105~K\cite{unoki67}.
This is due to a softening of the lowest frequency
triply degenerate zone-corner R-point vibrational
mode of lowest energy\cite{fleury68,shirane69}. 
In this mode, the Sr and Ti atoms remain
fixed, while the oxygen octahedra rotate about
one of the cubic axes passing through the Ti atom.
The sense of the rotation is opposite in adjacent cells in 
all directions as shown in Fig.~\ref{fig1}.
As the temperature continues to drop
below the AFD transition, there is a sharp rise
in the value of the static dielectric constant.
This increase appears to follow a Curie-Weiss law
consistent with an impending ferroelectric (FE) transition at
about 20~K.  With decreasing temperature, however,
the static dielectric constant saturates at a high value,
and the material does not become ferroelectric
even at the lowest temperatures\cite{sakudo71,viana94}. 
It has been suggested that the FE transition is
inhibited due to quantum paraelectric behavior,
in which the zero-point motion of the atoms suppresses
the long-range FE order\cite{muller79,muller91}.
It is thus of great interest to characterize the
structural instabilities of SrTiO$_3$.
The present work is concerned with the wavevector
dependence of structural instabilities in SrTiO$_3$,
obtained from first-principles calculations, and their
interpretation in light of previous similar calculations on KNbO$_3$.

\section{Methods}

 Our calculations are carried out utilizing the
LAPW formalism\cite{andersen75},
wherein the wave functions, charge density, and
potential all have dual representations.
Within non-overlapping muffin-tin spheres, these
quantities are expanded as numerical radial
functions times spherical harmonics,
while in the interstitual region, a plane wave
representation is used. This method has several
attractive features, such as employing an unbiased
representation in the interstitual region and the
ability to easily treat first-row and d-band atoms.
The major drawback presented by this method
is the use of position-dependent basis functions.
This introduces correction terms when computing
such quantities as the forces on the atoms\cite{yu91},
and thus complicates the computer programs. In
order to investigate all possible soft phonon modes
in a particular substance, we must map out the
dispersion curves throughout the Brillouin zone.  
This is accomplished quite easily using linear response
theory\cite{yu94,baroni87}.  In linear response
theory, the total variational energy is expanded to
second order as a function of the equilibrium
charge density and the first-order change in
the charge density:
\begin{equation}
  E_T = E^{(0)}[\rho^{(0)}] + \lambda E^{(1)}[\rho^{(0)}] +
        \lambda^2 E^{(2)}[\rho^{(1)}] + \cdots.
\end{equation}
Minimizing this energy with respect to the
first-order change in the charge density
yields a pair of equations which must be solved
self-consistently:
\begin{equation}
  \rho^{(1)} = \sum_{i=1}^N \psi_i^\ast \psi_i^{(1)},
\end{equation}
\begin{equation}
  \biggl ( H^{(0)} - \varepsilon_i^{(0)} \biggl ) \psi_i^{(1)} =
  - \biggl ( V_{eff}^{(1)}[\rho^{(1)}] \biggl )
  \psi_i^{(0)}.
\end{equation}
The equations are the first-order analogs of the
usual local density approximation (LDA)
equations, i.e. this is an exact theory; there are no
approximations beyond LDA.
\par
The dynamical matrix elements for a given
wavevector $\vec{{\bf q}}$ are found via the
first-order forces.  The forces on all the atoms
are calculated for at most 3N selected atomic
displacements at each wavevector, where N is the
number of atoms in the primitive cell.  To map the
phonon dispersion curves throughout the Brillouin
zone, we compute the dynamical matrix on a
uniform grid of q-points.  The short-range real-space
force constants are then obtained by Fourier transform, after
subtracting out the analytic long-range dipolar contribution.
By inverse Fourier transform, adding in the long-range part,
we may compute the dynamical matrix at an arbitrary
q-point.  For the SrTiO$_3$ calculations, we have
used a 6$\times$6$\times$6 mesh of q-points. Exploiting
symmetry, we had to compute the dynamical matrix
at only 20 q-points in the irreducible BZ wedge.
Typically, at each q-point, 3N=15 self-consistent
calculations were performed.
\par
  The electronic contribution to the static
dielectric constant, $\varepsilon_\infty$,
is determined from the induced potential:
\begin{equation}
  \hat{{\bf q}} \cdot \epsilon_{\infty} \cdot \hat{{\bf q}}
  = \lim_{ \bf q \rightarrow 0} ~ \biggl [ 1- \frac{V_{ind}
  (\vec{{\bf q }})}{V_{total}(\vec{{\bf q }}) } \biggl ].
\end{equation}
The macroscopic polarization at long wavelength is given by the first
order change in the charge density
\begin{equation}
  P = \lim_{\vec{\bf q} \rightarrow 0} ~
      \frac{i}{q} ~\rho^{(1)}({\vec{\bf q}}),
\end{equation}
and from this we can compute the Born effective
charges
\begin{equation}
  Z_{\alpha,\beta}^\ast (j) = Z_j + \Omega ~ \partial{P_{\alpha}}
  / \partial{\tau_{\beta} (j)} ~ \bigg | _{\vec{{\bf E}} = 0}.
\end{equation}
Here, Z$_j$ is the ionic contribution, and the second term is
the unit cell volume times the derivative of the
polarization with respect to displacements
of the atoms at zero electric field.

\section{Results}

We first performed self-consistent total-energy
calculations in the cubic phase to determine
the equilibrium properties.  Table~\ref{table1}
lists the theoretical lattice parameters and bulk
moduli for the present work, along with those
of King-Smith and Vanderbilt \cite{kingsmith94},
as well as some experimental values.
Our calculated lattice parameter is only 0.4\% larger
than the experimental value, so we have used
the theoretical value in the following
calculations.  The electronic component of the static 
dielectric constant, $\varepsilon_\infty$, was found to be
6.63, which is about 28\% larger than an experimental
value of 5.18.
This kind of overestimation of $\varepsilon_\infty$
is typical of LDA calculations\cite{gironcoli89,wang96}.

\par
Table~\ref{table2} lists the Born effective charges
(Z$^{\ast}$) for each type of atom.  For Sr and Ti, the 
values are isotropic, while for the O atoms, there
are two distinctive values---for displacements along and
perpendicular to the Ti-O bonds, labeled $\parallel$ and
$\perp$, respectively.  Also listed are the
results of calculations by Zhong, King-Smith,
and Vanderbilt\cite{zhong94}.  Respective theoretical 
lattice parameters were used for the calculations.
Note the anomalously large values for Z$^\ast$(Ti) and
Z$^\ast$(O)$_{\parallel}$ due to strong covalent
interactions between these atoms\cite{cohen90,cohen92}.
We find that our results are generally in good
agreement with those of Zhong {\it{et al.}}\cite{zhong94}.

\par
Table~\ref{table3} compares our zone-center optic phonon frequencies
with planewave pseudopotential frozen phonon 
calculations and experimental data.
Both calculations find unstable transverse optic (TO)
modes at the $\Gamma$-point
with imaginary frequencies.
The longitudinal optic (LO) mode frequencies were obtained using
\begin{equation}
  D_{i \alpha, j \beta}^{LO} = D_{i \alpha, j \beta}^{TO} +
  \frac{4 \pi e^2}{\Omega \sqrt{{M_i M_j}}}
  \frac{{({\bf{Z_i^\ast \cdot \hat{q}}})}_\alpha ({\bf{Z_j^\ast
         \cdot \hat{q}}})_\beta}{\varepsilon_\infty},
\end{equation}
where {\it{D}}$^{TO}$ is the zone-center dynamical matrix without
macroscopic field, {\bf{Z$_i^\ast$}} and M$_i$ are the Born
effective charge tensor and mass for atom $i$, $\Omega$ is the volume
of the unit cell, $\alpha$, $\beta$ are Cartesian indices, and
${\bf{\hat{q}}}$ is a unit wavevector. The results of
Zhong, King-Smith, and Vanderbilt were determined
using $\varepsilon_\infty$=5.18 from experiment, whereas we
used our larger calculated value of 6.63.  To determine
what effect $\varepsilon_\infty$ has on the LO mode
eigenvalues, a second calculation was performed
using $\varepsilon_\infty$=5.18.  The highest LO mode
is most sensitive, changing from 751 to 832 cm${^-1}$,
while the other two are fairly insensitive.

\par
We present the calculated phonon dispersion
curves for cubic SrTiO$_3$ plotted along
high-symmetry directions in Fig.~\ref{fig2}.
The $\Gamma$X, $\Gamma$M, and $\Gamma$R lines correspond to the
$\langle$100$\rangle$, $\langle$110$\rangle$, and
$\langle$111$\rangle$ directions, respectively.
Imaginary frequencies are represented by negative values.
The character of the modes at the zone-center
and zone-boundaries has been labeled according to the
notation by Cowley\cite{cowley64}.
The form of some of our lower dispersion curves
for the $\Gamma$M and $\Gamma$R directions is roughly
in accordance with data reported by Shirane and
Yamada at 120~K\cite{shirane69}, including an apparent
softening of a mode at the R-point.  Similar behavior is
seen at the M-point, but the lowest zone-boundary phonon
mode M$_3$ was reported as being temperature-independent
through the AFD transition.

\par
Since we are mainly concerned with structural instabilities
and how they relate to the observed phase transitions, we will
focus on the portions of the dispersion curves which lie
below the $\omega$=0 dashed line in Fig.~\ref{fig2}.
Note the large phase space for unstable modes.
Our calculations indicate unstable modes at
the R-point (R$_{25}$), zone-center ($\Gamma_{15}$),
and M-point (M$_3$).  These instabilities are
of two types: FE ($\Gamma_{15}$),
and AFD (R$_{25}$ and M$_3$).  
In the $\Gamma_{15}$ TO mode, Ti atoms move parallel
to one of the Ti-O bonds, and the oxygen octahedra
move in the opposite direction.  This mode is 
responsible for the FE transitions in materials
like BaTiO$_3$ and KNbO$_3$\cite{comes68,dougherty92}.
The M$_3$ mode is nearly identical to the R$_{25}$
mode in Fig.~\ref{fig1} with one exception.
The rotation of the octahedra is in the $same$ sense
in neighboring cells along the vertical
c-axis, but remains opposite in the horizontally
adjacent cells.  The dispersion curves of Fig.~\ref{fig2}
show $|\omega(\Gamma_{15})|<|\omega(R_{25})|$,
suggesting that LDA may find a FE structure (rhombohedral?)
to be lower in energy than the AFD
tetragonal structure. Of course experimentally
there is no observed transition from AFD to FE
phase, although it is seen in the Monte Carlo
simulations of Zhong and Vanderbilt\cite{zhong95}.
Recently, they have included quantum fluctuation
effects into these simulations, and find the FE
transition completely suppressed\cite{zhong96,vanderbilt96}.

\par
The regions of instability in the BZ are better
visualized in Fig.~\ref{fig3}, in which
isosurfaces, corresponding to $\omega$=0, are shown
for the lowest unstable phonon modes.  The
cubic BZ, with $\Gamma$ at the center, is also
shown.  The inner isosurface is centered at 
the $\Gamma$-point, and can be visualized as
three interpenetrating ``cookies'', one perpendicular
to each Cartesian direction.  The region of FE-type
instability is interior to this first isosurface.
Between the first and the second isosurface,
which lies near the zone edges, all modes
are stable.  Thus unstable modes are present
along the entire R-M-R edge of the BZ.
In a repeated-zone scheme, this isosurface would appear as
narrow cylindrical tubes extending from R-M-R.

\par
We now discuss the current work in light of results for
similar calculations on cubic KNbO$_3$ by Yu and Krakauer\cite{yu95}.
The dispersion curves of KNbO$_3$ also reveal
a large region of instability for the lowest modes.
In contrast to SrTiO$_3$, a soft mode is present
at the X-point, and the lowest R-point mode is stable.
Although M-point instabilities exist for both materials,
the character of these modes is quite different.  In
KNbO$_3$ the motion of the atoms is the same as
for the zone-center $\Gamma_{15}$ mode with
the exception that two O atoms
perpedicular to the motion remain fixed, i.e. it
is essentially the ``ferroelectric'' soft mode.
In SrTiO$_3$, the motion is a rotation of the octahedra as
described above.  A similar $\omega$=0 isosurface plot 
for KNbO$_3$ shows that the FE instability extends from
the zone-center all the way to the zone-boundary within
planar slab-like regions.  An AFD instablility was not
present in the KNbO$_3$ calculations.  Notice that in SrTiO$_3$,
the large planar regions of FE instability have
shrunk down to three interpenetrating ``cookies'',
and the phase space corresponding to the FE character
is greatly reduced.

\par
Because the regions of instability in KNbO$_3$ reside
in slabs which extend across the entire BZ, any linear
combination of phonon modes within one of these regions
will also be unstable. In a real-space picture,
unstable atomic chain-like structures can
develop, where the Nb atoms are all
displaced in the same direction\cite{yu95,comes68}.
The width of the slab-like regions in KNbO$_3$
indicates the length of the shortest such chains
to be about 5 lattice parameters\cite{yu95}. We can
perform a similar qualitative analysis of the real-space
motion of the atoms in SrTiO$_3$ from its isosurface
plot.  Since there exists a continuous instability
extending from R-M-R, linear combinations of
phonon modes along this cylindrical tube-like region
will also be unstable.  The character of modes 
along the zone edges is one of rotating oxygen octahedra;
thus, an unstable thin disk or planar region could
form in real space wherein the octahedra are rotated
in opposite senses in neighboring cells, with the
surrounding regions remaining undisplaced.  From
the width of the cylinders containing the 
instability, we estimate the smallest radius 
of the real-space disks to be approximately
3--5 lattice parameters.
Because of the sensitive dependence of the soft
modes on volume, it is possible that these
cylinders may pinch off at the M-point at
slightly reduced volumes.  We have not investigated
this yet.

\section{Summary}

In summary, we performed first-principles linear
response calculations of the wavevector
dependence of structural instabilities for cubic
SrTiO$_3$.  In comparison with previous work on
KNbO$_3$, we notice three differences.  First,
calculations on KNbO$_3$ show a ferroelectric-type
instability only, while SrTiO$_3$
calculations exhibit both FE and AFD instabilities.
Second, the phase space of the two-dimensional
FE instability (corresponding to chains in real
space) is greatly reduced and no longer extends
to the  zone-edges as in KNbO$_3$, but is
restricted to the interior of a region likened
to three inter-penetrating ``cookies''. Third,
the AFD instability occupies a rather large
region of phase space extending along the entire
BZ edges in cylindrical tubes.  The essentially
1-D character of these tubes corresponds to
real-space planar instabilities characterized
by rotations of the oxygen octahedra. Future
calculations for SrTiO$_3$ in the tetragonal phase
(below the AFD transition) will be required in order
to investigate any instabilities---especially
ferroelectric instabilities---that persist.

\acknowledgments

This research was supported by Office of Naval Research grant
N00014-94-1-1044.  C.-Z. Wang was supported by National Science
Foundation Grant DMR-9404954.  Computations were carried out at the
Cornell Theory Center.


\begin{figure}
\caption[1]{Motion of the oxygen atoms in the zone-corner
         R$_{25}$ phonon mode.  The Ti atoms lie inside
         the octahedra and do not move.  The Sr atoms are also
         fixed but are not shown.  The octahedra rotate about
         the c-axis in opposite senses in all adjacent cells.}
\label{fig1}
\end{figure}

\begin{figure}
\caption[2]{Calculated phonon dispersion curves for cubic SrTiO$_3$ at
         the theoretical lattice constant.}
\label{fig2}
\end{figure}

\begin{figure}
\caption[3]{Zero-frequency isosurfaces of the lowest unstable phonon
         modes over the entire BZ.  The $\Gamma$-point is located at
         the center of the cube.  Unstable modes exist inside the
         central surface and along the full length of the zone-edges.}
\label{fig3}
\end{figure} 

\begin{table}
\caption[t1]{Equilibrium properties determined by calculations and
           experiment for SrTiO$_3$.}
  \vskip 6mm
  \begin{center}
    \begin{tabular}{lll}
          & Lattice parameter $a$ & Bulk modulus \\
          & (a.u.) & (GPa) \\
      \hline
      Present & 7.412 & 190 \\
      PW$^{a}$ & 7.303 & 200 \\
      Exp.$^{b}$ & 7.380 & ~ \\
      Exp.$^{c}$ & ~ & 179$\pm$6 \\
    \end{tabular}
  \end{center}
  
  $^{a}$Planewave ultrasoft-pseudopotenial calculation by King-Smith and Vanderbilt \cite{kingsmith94}.\\
  $^{b}$Wyckoff, {\it{Crystal Structures}}, vol. 2, 2nd ed. (1964). \\
  $^{c}$T. Mitsui {\it{et al.}}, Landolt-Bornstein Series, Group III, vol. 3
         (1969). \\
\label{table1}
\end{table}

\begin{table}
\caption[t2]{Born effective charges for cubic SrTiO$_3$.}
  \vskip 6mm
  \begin{center}
    \begin{tabular}{lll}
         & Present & PW-BP$^{a}$ \\
       \hline
         Lattice parameter (a.u.) & 7.412 & 7.303 \\
       \hline
         Z$^{\ast}$(Sr) & 2.55 & 2.54 \\
         Z$^{\ast}$(Ti) & 7.56 & 7.12 \\
         Z$^{\ast}$(O)$_{\parallel}$ & -5.92 & -5.66 \\
         Z$^{\ast}$(O)$_{\perp}$ & -2.12 & -2.00 \\
    \end{tabular}
  \end{center}
  $^{a}$Planewave Berry phase calculation by Zhong {\it{et al.}}
  \cite{zhong94}. \\
\label{table2}
\end{table}

\begin{table}
\caption[3]{Calculated zone-center optic phonon
           frequencies (in cm$^{-1}$) in cubic SrTiO$_3$.}
  \vskip 6mm
  \begin{center}
    \begin{tabular}{lll}
       Present & PW$^a$ & Exp.$^b$(300~K) \\
       \hline
       TO modes: & & \\
       100$i$ & ~41$i$ & Soft \\
       151  & 165 & 175 \\
       219  &     & IR-silent \\
       522  & 546 & 545 \\
       \hline
       LO modes: & & \\
       146$^A$~~~~~~146$^B$ & 158 & 171 \\
       439$^A$~~~~~~449$^B$ & 454 & 474 \\
       751$^A$~~~~~~832$^B$ & 829 & 795 \\
    \end{tabular}
  \end{center}
  $^a$Reference\cite{zhong94}. \\
  $^b$Reference\cite{servoin80}. \\
  $^A$Using Eq.(7) and the calculated value $\varepsilon_\infty$=6.63. \\
  $^B$Using Eq.(7) and $\varepsilon_\infty$=5.18 extracted
      from experiment. \\
\label{table3}
\end{table}


\begin{references}

\bibitem{yu94}
  R. Yu and H. Krakauer, Phys. Rev. B {\bf 49}, 4467 (1994).

\bibitem{yu95}
  R. Yu and H. Krakauer, Phys. Rev. Lett. {\bf 74}, 4067 (1995).

\bibitem{unoki67}
  H. Unoki and T. Sakudo, J. Phys. Soc. Japan {\bf 23}, 546 (1967).

\bibitem{fleury68}
  P. A. Fleury, J. F. Scott, and J. M. Worlock, Phys. Rev. Lett.
  {\bf 21}, 16 (1968).

\bibitem{shirane69}
  G. Shirane and Y. Yamada, Phys. Rev. {\bf 177}, 858 (1969).

\bibitem{sakudo71}
  T. Sakudo and H. Unoki, Phys. Rev. Lett. {\bf 26}, 851 (1971).

\bibitem{viana94}
  R. Viana, P. Lunkenheimer, J. Hemberger, R. B\"{o}hmer,
  and A. Loidl, Phys. Rev. B {\bf 50}, 601 (1994).

\bibitem{muller79}
  K. A. M\"{u}ller and H. Burkard, Phys. Rev. B {\bf 19}, 3593 (1979).

\bibitem{muller91}
  K. A. M\"{u}ller, W. Berlinger, and E. Tosatti, Z. Phys. B
  {\bf 84}, 277 (1991).

\bibitem{andersen75}
  O. K. Andersen, Phys. Rev. B {\bf 12}, 3060 (1975);
  E. Wimmer, H. Krakauer, M. Weinert, and A. J. Freeman,
  Phys. Rev. B {\bf 24}, 864 (1981);
  S.-H. Wei and H. Krakauer, Phys. Rev. Lett. {\bf 55}, 1200 (1985);
  D. J. Singh, {\it{Planewaves, Pseudopotentials and the LAPW Method}}
  (Kluwer Academic Publishers, Massachusetts, 1994).

\bibitem{yu91}
  R. Yu, D. Singh, and H. Krakauer, Phys. Rev. B {\bf 43}, 6411 (1991).

\bibitem{baroni87}
  S. Baroni, P. Giannozzi, and A. Testa, Phys. Rev. Lett.
  {\bf 59}, 2662 (1987).

\bibitem{kingsmith94}
  R. D. King-Smith and D. Vanderbilt, Phys. Rev. B {\bf 49}, 5828 (1994).

\bibitem{gironcoli89}
  S. de Gironcoli, S. Baroni, and R. Resta, Phys. Rev. Lett.
  {\bf 62}, 2853 (1989).

\bibitem{wang96}
  C. Z. Wang, R. Yu, and H. Krakauer, Phys. Rev. B {\bf 53}, 5430 (1996).

\bibitem{zhong94}
  W. Zhong, R. D. King-Smith, and D. Vanderbilt, Phys. Rev. Lett.
  {\bf 72}, 3618 (1994).

\bibitem{cohen90}
  R. E. Cohen and H. Krakauer, Phys. Rev. B {\bf 42}, 6416 (1990).

\bibitem{cohen92}
  R. E. Cohen, Nature (London) {\bf 358}, 136 (1992).

\bibitem{cowley64}
  R. A. Cowley, Phys. Rev. {\bf 134}, A981 (1964).

\bibitem{comes68}
  R. Comes, M. Lambert and A. Guiner, Solid State Commun.
  {\bf 6}, 715 (1968).

\bibitem{dougherty92}
  T. P. Dougherty {\it{et al.}}, Science {\bf 258}, 770 (1992).

\bibitem{zhong95}
  W. Zhong and D. Vanderbilt, Phys. Rev. Lett.
  {\bf 74}, 2587 (1995).

\bibitem{zhong96}
  W. Zhong and D. Vanderbilt, Phys. Rev. B (in press).

\bibitem{vanderbilt96}
  D. Vanderbilt, W. Zhong, J. Padilla, and A. Garcia 
  (this volume).

\bibitem{servoin80}
  J. L. Servoin, Y. Luspin, and F. Gervais, Phys. Rev. B
  {\bf 22}, 5501 (1980).

\end{references}
\end{document}